# Advancing Drug Development Through Strategic Cell Line and Compound Selection Using Drug Response Profiles

Abbi Abdel-Rehim[1,*], Emma Tate[2], Larisa N. Soldatova[3], and Ross D. King[1,4,5,6]

[1]Department of Chemical Engineering and Biotechnology, University of Cambridge, Cambridge, UK.

[2]Arctoris Ltd, Oxford, UK.

[3]Department of Computing, University of London, London, UK.

[4]Department of Biology and Biological Engineering, Chalmers University of Technology, Gothenburg, Sweden.

[5]Department of Computer Science and Engineering, Chalmers University of Technology, Gothenburg, Sweden.

[6]The Alan Turing Institute, London, UK.

*Corresponding author: aar52@cam.ac.uk

**Abstract**

Early identification of sensitive cancer cell lines is essential for accelerating biomarker discovery and elucidating drug mechanism of action. Given the efficiency and low cost of small-scale drug screens relative to extensive omics profiling, we compared drug-response panel (DRP) descriptors against omics features for predictive capacity using gradient boosting tree models across the GDSC and CCLE drug response datasets. DRP descriptors consistently outperformed omics data across key performance metrics, with variable performance across different drugs. Using complementary explainability approaches, we confirmed known MAPK-inhibitor sensitivity signatures, and identified novel potential biomarker candidates for MEK1/2 and BTK/MNK inhibitors. Lastly, to demonstrate the utility of this approach in distinguishing phenotypes, we applied our models to the breast cancer line MCF7 versus the non-tumorigenic MCF10A, and successfully identified compounds that selectively inhibit MCF7 while sparing the non-tumorigenic MCF10A. This methodology, developed using focused drug and cell line panels, supports early-stage drug development by facilitating rational cell line selection and compound prioritisation, enabling more efficient biomarker identification and candidate assessment.

**Introduction**

The development of cancer therapies often begins by identifying a specific molecular target, followed by the optimization of a drug to achieve strong affinity and specificity. Initial efficacy is typically evaluated in vitro using cell culture models. However, the inherent heterogeneity of cancer necessitates testing across multiple cell lines to confirm the compound's intended toxicity. At this early stage, testing is usually limited to pre-selected cell lines based on target validation and tissue type. Given the extensive diversity of available cell lines, and the resource-intensive nature of these experiments, it is impractical to evaluate every candidate compound across a comprehensive range of cell line models. Strategies that optimise early cell line selection could accelerate discovery and improve downstream clinical success rates.

Cell line assays provide crucial insights into a drug's mode of action, which is particularly important for targeted therapies [1, 2]. These insights can also reveal biomarkers indicative of drug response, which enhances the probability of success during the challenging transitioning from *in vitro* models to *in vivo* studies and clinical trials [3, 4]. While tissue of origin shapes the genomic landscape, it often proves too coarse to effectively guide targeted therapies. For any tissue type, drugs appear to be effective only in subpopulations [5], even where a genetic marker has been identified [6, 7, 8, 9]. Thus, early and efficient identification of molecular determinants of response is critical. To achieve this, a key strategy is to predict and select an optimised set of cell lines that are highly sensitive to a compound. These pre-selected, sensitive cell lines can then be used in focused downstream mechanistic studies to extract relevant biological information and robust biomarkers.

Recent efforts have turned to omics profiling to guide drug sensitivity predictions; however, these approaches often require large sample sizes, suffer from limited generalizability across drugs, and can be challenging to interpret [10, 11, 12]. Emerging descriptor-based models, such as drug response panels (DRPs), offer a promising alternative by capturing phenotypic drug sensitivity profiles without relying exclusively on molecular data [13, 14, 15]. Yet, their comparative advantages and integration into early-stage drug discovery workflows remain underexplored. In this study, we develop predictive models for drug activity in cell lines, systematically comparing omics-based features to drug response panel (DRP) descriptors using gradient boosting trees models (Figure 1).

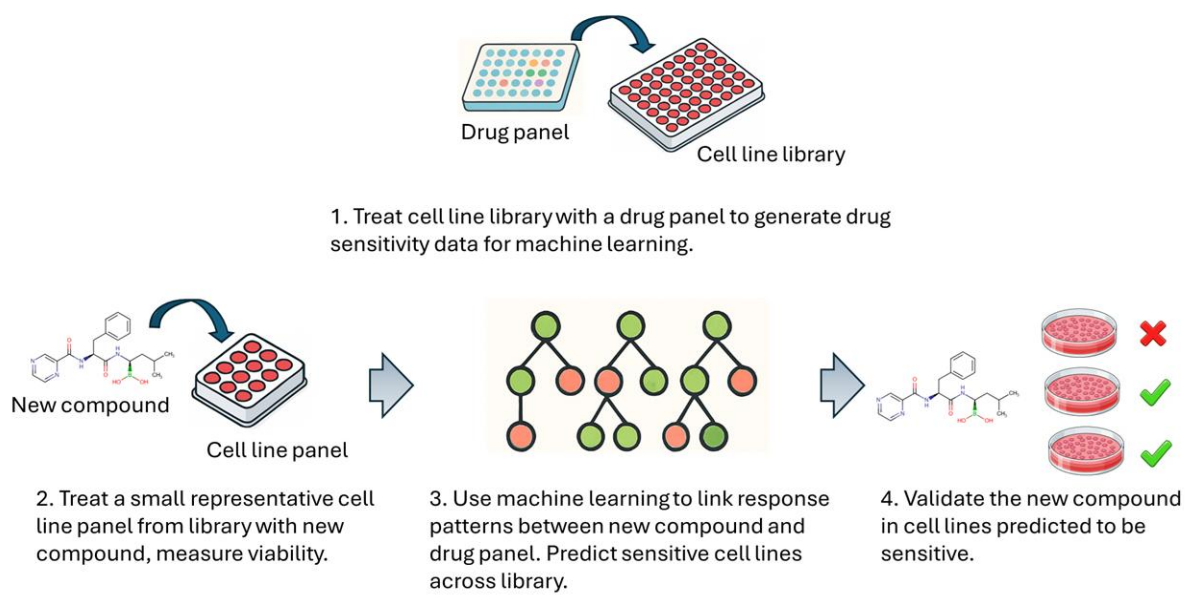

*Figure 1. Concept of DRP-based prediction for new compound activity. Cell lines are characterised by their response to a fixed drug panel. New compounds are tested in a subset of these cell lines. Machine learning is used to learn relationships between the new drug's responses and those of the drugs in the panel. The generated model is then used to predict the activity of the new drug in other cell lines that have been screened with the drug panel but not yet tested with the new drug.*

We show that DRP descriptors outperform traditional omics in predicting drug sensitivity using Genomics of Drug Sensitivity in Cancer (GDSC) and Cancer Cell Line Encyclopedia (CCLE) datasets. We also perform a 10-fold cross validation using GDSC for benchmarking purposes, and study how the performance differs when considering each drug individually instead of all datapoints together (Figure 2). Using complementary explainability approaches, including feature-response correlations, random forest feature importance, and SHAP (SHapley Additive exPlanations) analysis, we confirmed known MAPK-inhibitor sensitivity signatures and uncovered novel potential candidates, including regulators of Rho GTPases, Ras/ERK and BCL-2 pathways. These findings highlight the ability of even moderately sized cell line panels to yield biologically meaningful insights.

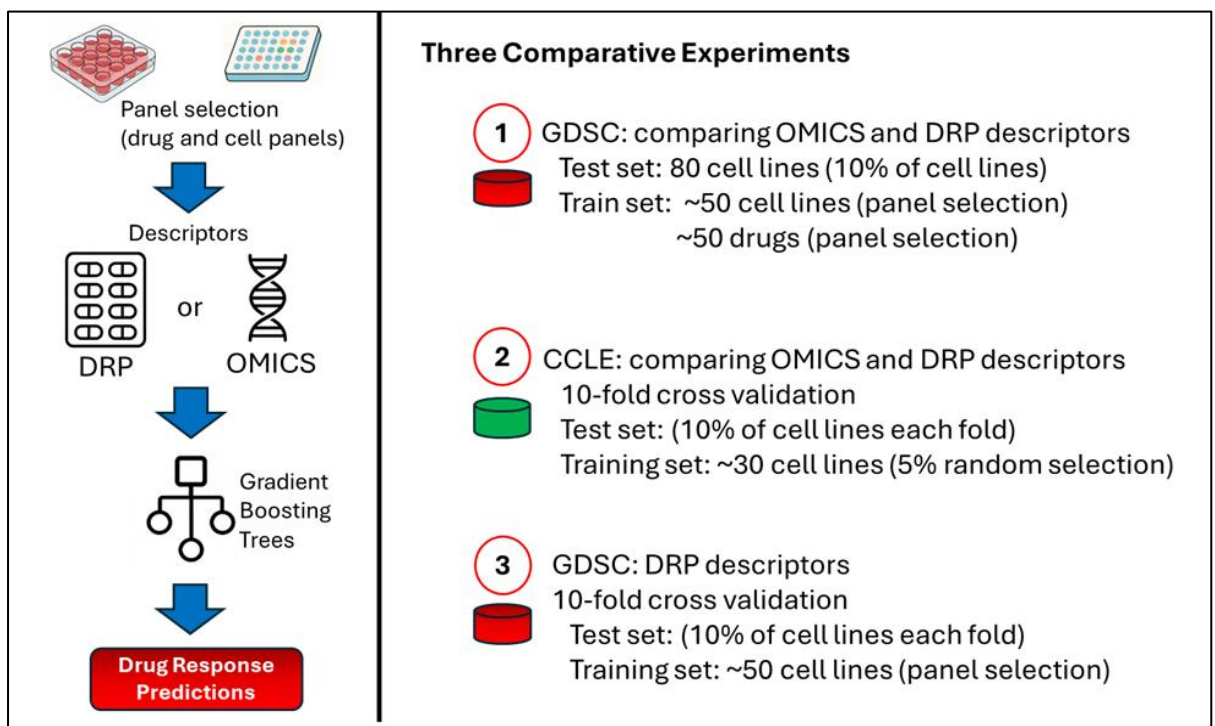

*Figure 2. Three benchmarking experiments. (1) Comparing performance of the omics and DRP descriptors on the GDSC dataset using a dedicated test set. (2) Comparing performance of the descriptors on the CCLE dataset using cross validation. (3). Comparing performance of DRP descriptors when considering all datapoints in the GDSC dataset or each drug individually.*

Lastly, we tested whether DRP-based machine learning models could guide prospective compound selection to distinguish between cancerous and non-cancerous epithelial lines. Focusing on breast epithelial MCF7 versus non-tumorigenic MCF10A cells, our models successfully nominated compounds with divergent, lineage-specific effects, validated experimentally.

Altogether, our work demonstrates that predictive modelling based on DRP descriptors offers a robust and efficient strategy for early-stage drug discovery. By enabling rational cell line selection, uncovering biologically meaningful response determinants through explainability analyses of such selections, and prospectively nominating compounds that discriminate between malignant and non-malignant tissues, our approach provides a scalable framework to enhance compound prioritisation, guide biomarker discovery, and improve translational relevance early in the drug development pipeline.

**Methods**

**Datasets:**
GDSC 1 and 2 drug response data ($IC_{50}$ values - https://www.cancerrxgene.org/), CCLE drug response data ($IC_{50}$ values - https://depmap.org/), mRNA expression profiles (TPM values - https://cellmodelpassports.sanger.ac.uk/).

All datasets were processed and consolidated. The GDSC datasets contained overlapping drugs, and some drug and cell names varied between datasets. We manually aligned the cell lines with differing names and compared drugs using their MACCS fingerprints, keeping redundancy out. Non-small molecule drugs and those without a structure for MACCS fingerprinting were excluded. The mRNA data was log2-transformed, followed by dimensionality reduction using scikit-learn's (version 1.2.0) implementation of PCA, retaining components that accounted for 90% of the variance in the dataset. PCA was not applied for the explainability studies.

To promote reproducibility and benchmarking, we have uploaded the processed GDSC and CCLE datasets containing $-1*\log10(IC_{50})$ values, along with clearly labelled cell and drug names. The mRNA expression datasets containing log2(TPM) values, along with PCA-transformed data used in model building, are also available.

**MACCS fingerprints:**
MACCS molecular fingerprints were generated using RDKit (version 2022.09.3), based on molecular SMILES codes from PubChem, matched to the drug names used in the datasets.

**Panel Selection:**
Samples (cell lines or drugs) exhibiting low variation (bottom 50%) are excluded. A Pearson correlation matrix is constructed using the remaining samples, which are ordered based on their median correlation values. The process begins by adding the sample with the lowest correlation to the panel, after which any samples with correlations above a user-defined threshold are excluded. This procedure continues down the list until no samples remain. The correlation threshold is adjustable, allowing for flexibility in panel stringency. This approach ensures that only the top 50% of varying samples, with minimal correlation to others, are included in the panel.

**Machine Learning:**
We employed scikit-learn implementation of Gradient Boosting Trees (version 1.2.0) for drug activity prediction, selected for its strong performance in this context [16]. A variety of tree numbers and depths were tested (Table S1) across 58 randomly selected drugs from the GDSC dataset in 5 independent experiments per drug. The most basic model was chosen, as additional depth and trees did not significantly improve predictions.

**Explainability:**
For the explainability analysis, we employed Pearson correlation to assess the relationship between gene expression and cell line sensitivity to three drugs: Refametinib (MEK1/2 inhibitor), PD0325901 (MEK1/2 inhibitor), and QL-X-138 (BTK/NMK inhibitor). Cell lines were selected based on their predicted sensitivity to these drugs (n=3). For each experiment and drug, tree-based models were built using the RandomForestRegressor implementation from scikit-learn (200 trees, default settings). The 500 genes with the highest absolute correlation to the drug sensitivity were used to build the models. Feature importances were extracted using the built-in feature_importances_ function. Additionally, we performed SHAP (SHapley Additive exPlanations) analysis using the Python shap package, datapoints in the bee-swarm plot represents all data not used in training.

**Predicting drug responses for experimental validation:**
To predict responses in MCF10A and MCF7, the drug response data for MCF10A, serving as the drug response panel, was sourced from ChEMBL. Many of these drugs had been measured in few or no other cell lines. To address this, QSAR models were built based on each cell line in the NCI60 dataset, using MACCS fingerprints to predict the activity ($-\log(IC_{50})$ values) of drugs that had been measured in MCF10A. Given the large dataset size, two hyperparameters for the GradientBoostingRegressor were increased (100 trees, tree depth of 4).

Using these predictions, a drug response panel was established and employed to train a predictive DRP-based model. Except for MCF10A, all cell lines we've used has been measured with the GDSC drug library. Hence, the experimental data from GDSC was used to predict drug activities for both MCF10A and MCF7. Although MCF7 had been measured with many of the compounds in the library, we used predicted activities to demonstrate our method's predictive capabilities for both cell lines.

**Drug Screening:**

Drug library. 190 drugs, see Table XZ for details.

Cell Culture: Cell viability was assessed using the MCF7 and MCF10a cell lines. Cells were maintained in relevant media (see tables 1-2) throughout the study, in an incubator set to 37°C and 5% CO2, with media changes every 2-3 days. Cells were passaged when reaching ~80% confluency (typically every 3-4 days) using 1X TrypLE Select (MCF7) and PromoCell Detach Kit (MCF10a).

Compound plate preparation: Vehicle control (DMSO) or specified compound combinations was added to empty wells of a white-walled 384 well plate in a randomized fashion (one plate per cell line) using the D300e digital dispenser. Well volumes were normalized to 404nL with a consistent vehicle concentration of DMSO per well (0.5% (v/v) final assay DMSO concentration). Plates were sealed and stored at -4°C prior to use.

Viability Assay: Cells were seeded into compound-containing plates using the Dragonfly 10-Channel Dispenser (see section 2.2) at a density of $5.0 \times 10^3$ cells/cm2 (300 cells/well) in 40 µL of relevant growth medium. Plates were incubated for 48 h in a humidified incubator set to 37°C and 5% CO2. At the end of the incubation period, cells were equilibrated to 30°C and 40 µL of Cell Titer-Glo® 3D cell viability assay reagent was added (Promega; Cat #G9682). Plates were centrifuged briefly and incubated protected from light for 30 min at 30 °C. Luminescence signal was measured using a CLARIOstar Plus microplate reader (BMG Labtech).

Data Analysis: Raw relative luminescence unit (RLU) data were derandomized prior to background subtraction. Viability was determined as a percentage of vehicle control (DMSO only). % Viability = (RLU - Mean Positive control) / (Mean Negative control - Mean Positive control) *100.

# Results

## Drug response prediction and cell line selection

Predicting cell lines responses to compounds has the potential to play an important role in drug discovery, offering insights into drug efficacy and mechanism of action. Omics-based approaches have been widely applied in this context, typically relying on complex neural network architectures [12, 17, 18, 19, 20].

In this study, we explored the utility of drug response panels to characterise cell lines, and used these as features to generate predictive machine learning models (Figure 1). Using the GDSC dataset, we compared the DRP-based descriptors to omics features in the form of PCA-transformed mRNA expression profiles. Our choice of gradient boosting trees for benchmarking is based on their demonstrated efficacy in this domain [21-23]. Additionally, tree-based models tend to be more interpretable than complex neural networks, allowing for clearer feature importance analysis, which is useful in linking omics features to drug response. Drug panel and cell line panel selections used to generate the models were performed on training data prior to model training. Initial experiments including mutation and CNV descriptors with mRNA profiles did not yield notable performance improvements (not published), so we applied the mRNA descriptors alone. Dimensionality reduction via PCA, retaining 90% of variance, reduced computation with minimal impact on performance.

First, we randomly allocated patient derived cell lines to training (80%), validation (10%) and test-sets (10%). To determine the optimal settings for our learner, we randomly selected 58 drugs (20% of the total) and evaluated gradient boosting trees with two hyperparameters: tree depth (2,3 and 4) and the number of trees (50 and 100). For each of the five experiments. A drug panel selection was performed using the training-set resulting in a panel size of ~40-50 drugs. We used a randomly selected subset of 250 cell lines from the same training set to train the models. Based on previous findings, increasing the number of cell lines beyond this number won't provide notable improvements in performance [14]. We did not observe any significant performance differences across the settings (Table S1), therefore we selected the lower settings of 50 trees and a depth of 2 as the default configuration.

Using this optimised model, we evaluated performance on a test-set comprising all drugs in the GDSC dataset. While each drug was predicted individually as illustrated in figure 1, evaluation was carried out across all drug-cell line pairs in the test set simultaneously. Panel selection settings meant that approximately ~50 drugs were used in the drug panel and ~50 cell lines were used in the cell line panel for training the models. Due to the robustness of the panel selections causing close to identical results, a random subset of 70% of the training set cell lines was used for panel selections in each experiment. Both descriptor sets achieved high correlations (pearsonR and spearmanR) and low prediction errors (MSE, RMSE, MAE) relative to true activity values (Table 1). Notably, the DRP descriptors consistently and significantly outperformed the mRNA expression profiles. We also attempt these experiments using cell line selection that includes ~10, ~30 patients and ~100 patients for comparison (Table S2), these results show small but significant gains in performance at every increase.

|      | PEARSON | SPEARMAN | MSE | RMSE | MAE |
|------|---------|----------|-----|------|-----|
| **DRP** | 0.904±0.002 | 0.877±0.002 | 0.246±0.004 | 0.496±0.004 | 0.370±0.003 |
| **OMICS** | 0.834±0.002 | 0.793±0.004 | 0.434±0.008 | 0.658±0.006 | 0.503±0.005 |

*Table 1. Comparing performance between mRNA expression profiles and drug response profiles in predicting drug response for all drugs across 80 cell lines from the GDSC dataset, results are declared along with standard deviations resulting from 5 independent experiments. Statistical significance of differences in performance between descriptor sets was assessed using a paired t-test on RMSE (p< 0.001).*

Despite only including a small number of measurements of each target drug in the training data, our methodology approximates a leave-drug-out scenario, avoiding the overoptimism seen in random splits that include an abundance of all drugs in training. Even using the smaller cell line panel of 10 cell lines (Table S2), the DRP consistently outperformed published methods shown in Table **S3**. Interestingly our application with omics features also outperformed the published methods, highlighting the utility of limited cell line panels to boost response predictions.

We also tested our approach on a second dataset (CCLE), which includes 24 drugs tested across more than 500 cell lines. To enable comparison with previous published methods, we used 10-fold cross-validation. For every fold, each drug was predicted separately. For each drug, 5% of the training samples (~20 cell lines) were randomly selected for training. The generated models were then tested on the held-out set (comprising 10% of all cell lines). The use of random selection rather than response-based cell line panel selection, was motivated by the small number of drugs in the dataset. This small selection of samples during training reflects the intended use case: predicting drug responses across a full library of cell lines after testing the compound in only a small panel. This experiment showed that DRP descriptors again outperformed omics-based features (Table 2).

|       | PEARSON     | SPEARMAN    | MSE         | RMSE        | MAE         |
|-------|-------------|-------------|-------------|-------------|-------------|
| DRP   | 0.861±0.017 | 0.710±0.024 | 0.198±0.032 | 0.443±0.035 | 0.247±0.019 |
| OMICS | 0.784±0.022 | 0.587±0.013 | 0.302±0.033 | 0.547±0.031 | 0.328±0.015 |

*Table 2. Experiments comparing DRP and omics features on the CCLE dataset using 10-fold cross validation with a limited cell line panel derived from the training folds. Statistical significance of differences in performance between descriptor sets was assessed using a paired t-test on RMSE ($p < 0.001$).*

The results shown thus far are based on the performance for all predictions across all compounds. However, it is important to understand the variability in performance of a method when applied to individual compounds. To do this, we repeated our analysis on the larger GDSC dataset using 10-fold cross-validation. For each fold, a small subset of cell lines was selected from the training panel using our panel selection method (Figure 2). Performance was then evaluated both globally (all drugs combined as done above) and at the individual drug level (Table 3). Here it became evident that while performance of the methodology was strong overall, there was substantial variation at the individual drug level, with some drugs showing high predictability and others low (Appendix A). Notably, while correlation scores varied substantially between the two approaches, error metrics remained similar, highlighting the limitations of traditional error-based assessments in capturing model utility.

|          | PEARSON     | SPEARMAN    | MSE         | RMSE        | MAE         |
|----------|-------------|-------------|-------------|-------------|-------------|
| OVERALL  | 0.912±0.002 | 0.887±0.001 | 0.223±0.004 | 0.472±0.005 | 0.352±0.003 |
| PER-DRUG | 0.624±0.149 | 0.612±0.147 | 0.453±0.126 | 0.223±0.142 | 0.352±0.103 |

*Table 3. Comparison of DRP-based model performance on the GDSC dataset using two evaluation approaches: overall (all data points combined) and per-drug (evaluating each drug individually). Results are reported as averages ± standard deviations from five independent experiments.*

To further investigate the factors driving predictive performance across different drugs, we focused our analysis on selective drugs, those effective in fewer than 25% of cell lines in the GDSC dataset. We aimed to determine if strong model performance was primarily driven by direct target similarity or by functional relationships captured by the DRP descriptors. For this analysis we deployed our approach with DRP descriptors on a held-out set of 160 cell lines (~20% of the dataset). Reformulating the problem as a classification task (defining sensitivity as $IC_{50} <= 1\mu M$), we identified 10 drugs whose models successfully recovered at least 75% of all true positives, while requiring testing fewer than 30% of the 160 cell lines.

These-high performing models covered several kinase inhibitors, many of which had functionally related compounds in the training set, suggesting that shared target information contributed to their successful prediction (Table S4). However, for certain drugs such as NG-25 (targeting TAK1 and M4K), the top three compounds contributing to model performance had similarly weighted feature importance scores but did not share known primary targets with NG-25. This result points to a more complex target interplay or potentially unknown shared mechanisms of action.

We extended this mechanistic investigation to the most influential DRP descriptors driving prediction performance in the top 20 high-performing models. In 10 cases, the top-ranked predictive compound in the panel targeted a different molecule(s) than the drug being predicted. Critically, in 8 of these cases, the top three most influential drugs all had different known targets (Table S4). These observations strongly suggest that indirect biological relationships or polypharmacology, rather than simple target overlap, may underline the robust predictive power in many of the DRP-based models. Crucially, this capability allows researchers to identify novel functional connections between seemingly unrelated compounds in the DRP. This provides a mechanism-agnostic context for the compound being tested, offering unexpected leads for combination therapies and valuable new hypotheses regarding the compound's mode of action or potential off-target effects.

**Explainability**

Cell lines offer a unique advantage in early drug development by revealing molecular environments facilitating drug sensitivity. While the primary goal of our methodology is the efficient selection of cell lines (a task where DRPs proved superior), the ultimate purpose of this selection is to extract useful biological information, a task where omics data remains essential.

We therefore demonstrate the utility of our methodology by using the DRP predicted sensitivities to rationally construct an enriched cell line panel designed for biomarker discovery. This panel combines the initially selected lines with the top and bottom 10 predicted responders (~55-60 cell lines in total), creating a high contrast training set for mechanistic analysis. Using this refined panel, we trained random forest regression models to predict responses based on mRNA expression for three high performing drugs in the GDSC dataset: Refametinib and PD0325901 (MEK1/2 inhibitors), and QL-X-138 (BTK/NMK inhibitor). To identify key genes associated with drug sensitivity, we analysed the top correlating genes and assessed their importance using feature importance extraction and SHAP analysis.

Consistent with literature, the expression levels of the direct targets (MEK1/2) did not predict cell-line sensitivity to Refametinib. This reinforces that functional pathway state and regulatory elements are the critical determinants of drug efficacy. Accordingly, MAPK regulators ETV5, SPRY2 and TLR4 were amongst the most important predictors (Figure 2), confirming reported MAPK-inhibitor sensitivity signatures and implying pathway rewiring [24-26]. Excluding low GKAP1-expressing cells further enriched for sensitivity, highlighting GKAP1 as a potential novel biomarker with limited prior characterization in cancer [27]. A similar signature emerged for PD0325901, another MEK1/2 inhibitor, validating our methodology.

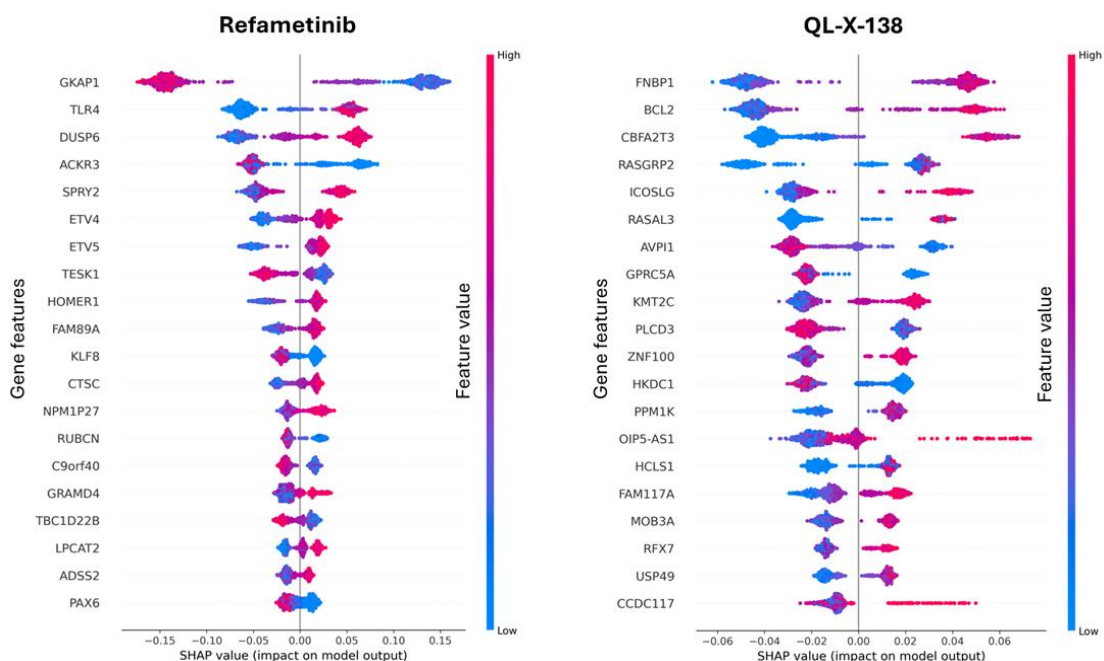

Figure 2. SHAP bee swarm plot showing feature importances from a random forest model trained to predict cell line sensitivity to Refametinib and QL-X-138 using mRNA expression levels of the top 500 genes correlated with sensitivity. Results from the first of three replicate experiments are shown.

For QL-X-138 (BTK/MNK dual inhibitor), the model showed more variability across experiments, but consistently highlighted markers such as WAS, RASAL3 and SEPTIN6 (positive correlation) and PLCD3 and AVPI1 (negative correlation) (Figure 2). Notably, BCL2 frequently appeared among the top features; this finding is biologically relevant as BTK inhibition has been linked to increased BCL-2 dependence, further supporting the validity of our signature discovery. Additionally, the role of RASAL3's connection to RAS/ERK pathway hints at novel pathway dependencies and potential crosstalk relevant to this dual inhibitor.

Finally, to validate the integrity of our approach, we tested whether the compact, performance driven panels selected by our methodology captured the broader biology derived from larger datasets. Models built on substantially larger random subsets (70% of cell lines) shared a high overlap of top SHAP derived features (Figure S1) for MEK1/2 inhibitors Refametinib and PD0325901, confirming that smaller, performance-driven panels can effectively capture the broader biology of drug response. The same experiment for QL-X-138 showed less feature overlap. This result is likely explained by the inherent variability observed previously, possibly amplified by highly correlated mRNA expression features.

**Validating DRP Versatility: Compound Prioritization for Selective Toxicity**
Our previous analysis demonstrated the utility of our DRP methodology for a core objective in drug development: biomarker discovery and cell line prioritization for a candidate drug. To further demonstrate the methodology's versatility, we now address a critical, complementary goal in the drug discovery pipeline: high-throughput compound screening for selective toxicity (Figure 3).

While our primary work focused on predicting the response of many cell lines to a single drug, this experiment inverts the problem to predict the response of two specific cell lines (MCF7 and

MCF10A) to a fixed library of 190 drugs. Our objective was to efficiently prioritise compounds that targeted the epithelial breast cancer cell line MCF7 while sparing the non-tumorigenic breast epithelial control line MCF10A, simulating a scenario where selective toxicity is critical for therapeutic relevance.

A key methodological challenge was creating a consistent DRP feature set suitable for predicting MCF7 and MCF10A responses. Although ChEMBL provided activity data for a significant number of compounds tested on MCF10A, the overlap of these compounds with the broader cell line panel was limited. To overcome this, we first harmonized the drug response profile by using standard QSAR models with MACCS fingerprints to predict the activities of 223 MCF10A-measured compounds across all other cell lines. We then constructed a consistent drug panel: MCF10A retained its measured activities, while the other cell lines were assigned their corresponding QSAR predictions. This process ensured all cell lines shared a common set of compounds, making these DRP features suitable for our modeling framework. Using this aligned drug panel, we applied our DRP approach to predict responses from the 190-compound library in both MCF10A and MCF7. Although experimental data for this compound library were available for MCF7, we used predicted values for both lines to simulate a realistic early-screening scenario where such data would be unavailable, relying on the reliable, experimentally tested data from the broader training set.

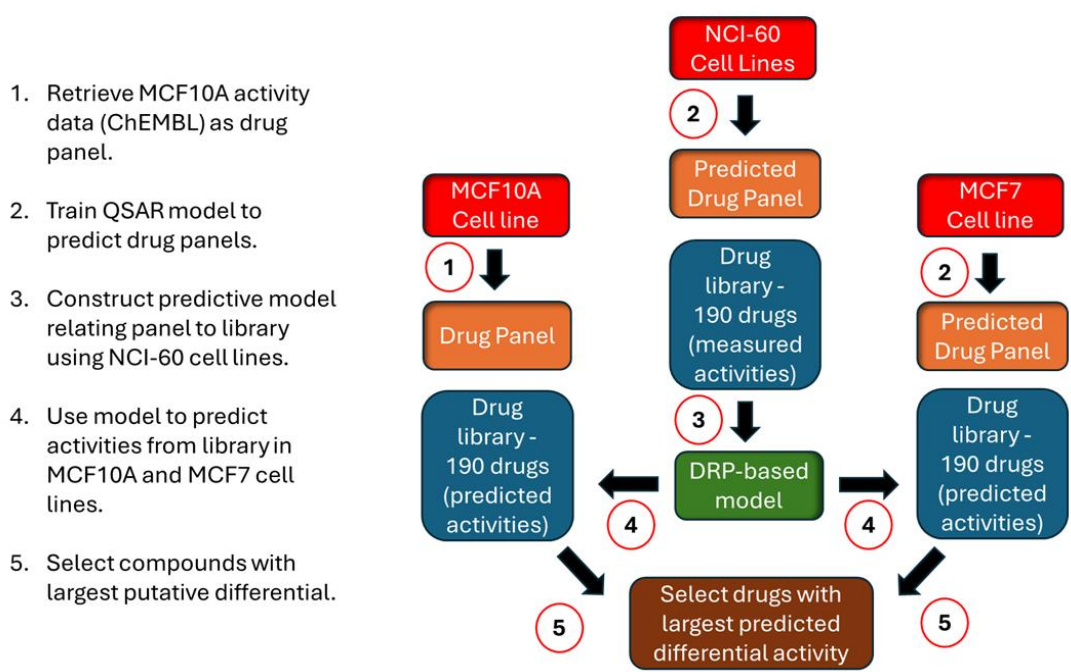

*Figure 3. Workflow for generating putative MCF7 and MCF10A activity profiles from a 190-compound drug library. 1) MCF10A activity data were retrieved from ChEMBL to define a drug panel (223 compounds). 2) Due to limited overlap with the NCI-60 dataset, QSAR models were trained to predict responses to these panel drugs across NCI-60 cell lines. This step generated a harmonized drug response profile (DRP). 3) DRP-based machine learning models were then constructed using the predicted panel responses and measured library activity in NCI-60. 4) These models were subsequently used to predict responses to the 190-compound library for the target cell lines MCF10A and MCF7. 5) Finally, four compounds with the highest predicted differential activity (max($IC50_{MCF7}$ - $IC50_{MCF10A}$)), were selected for experimental validation.*

From the predicted responses, we prioritised four compounds expected to have the largest differential effect (Max$_{\Delta viability}$), impacting MCF7 over MCF10A. We also included two additional compounds predicted to be more toxic towards MCF10A, to serve as negative controls. However, the largest predicted differences were modest (less than an order of magnitude), suggesting similar responses in both cell lines. All six compounds were tested experimentally across a fixed concentration range (2.5 µM 0.25 µM, 0.025 µM and 0.0025 µM).

The experimental results successfully identified two selective agents: Pevonedistat and BI 2536, two of the four prioritized compounds, which exhibited strong selectivity (Table 4). Both drugs achieved IC$_{50}$ values below 100 nM in MCF7. The selectivity was pronounced between the cell lines: in the control line MCF10A, BI 2536 had an IC$_{50}$ closer to 1 µM and Pevonedistat exceeded 2.5 µM. Furthermore, the model correctly identified the non-toxic nature of the two negative controls: neither Pemetrexed nor AZD6482 produced IC$_{50}$ values in either cell line, although MCF10A was slightly more sensitive to AZD6482 at the highest tested concentration.

| Drug | Predicted Preference (Δ10$^n$) | Measured Preference | MCF7 IC$_{50}$ Reached | MCF10 IC$_{50}$ Reached |
| --- | --- | --- | --- | --- |
| **Pevonedistat** | MCF7 (>2) | MCF7 | Y | N |
| Gemcitabine | MCF7 (>2) | MCF10 | N | Y |
| Cytarabine | MCF7 (>2) | None | N | N |
| **BI 2536** | MCF7 (>2) | MCF7 | Y | Y |
| **Pemetrexed** | None (<1) | None | N | N |
| **AZD 6482** | None (<1) | None | N | N |

*Table 4. Differential drug response predictions in MCF7 and MCF10A. Predicted and measured sensitivities for six tested drugs are shown, including predicted differential activity (Δ10$^n$), experimental outcomes (n=3), and whether IC$_{50}$ was reached at ≤2.5 µM. Bold indicates predictions consistent with experimental results.*

However, the model exhibited some misclassifications among the prioritized compounds: **Gemcitabine**, which had been predicted to favour MCF7, showed higher potency in MCF10A (IC$_{50}$ <0.25 µM), and Cytarabine showed little differential activity, with both lines maintaining above 70% viability at the highest tested concentration of 2.5 µM. Full experimental results (cell viability) are provided in Table S5.

**Discussion**
Our study demonstrates the viability of using drug response profiles (DPR) descriptors with machine learning to predict cell line sensitivity across a range of compounds. By leveraging diverse datasets, and benchmarking our models against published methods, we demonstrated the power of DRP-based features. The DRP approach is particularly valuable because it provides orthogonal information to static omics features. The resultant response matrices implicitly encode the complex biological relationships and functional dynamics influencing drug sensitivity, effectively capturing pathway flux and compensatory mechanisms. This functional perspective highlights the practicality and efficiency of DRP-based modelling. Unlike omics-driven methods that require extensive, multi-layer molecular profiling, DRPs enable accurate predictions even in low-data scenarios. This flexibility offers dual utility in the discovery pipeline. First, new cell lines can be quickly profiled against a limited panel of compounds to unlock predictive insights across a multitude of others, and second, novel drugs need only be tested in a small cell line panel to efficiently assess their broader activity landscape.

In addition, we demonstrated that DRP-guided cell line selection can support mechanistic interpretation when paired with omics-based models and explainability techniques like SHAP. For MEK1/2 inhibitors we recovered known MAPK pathway regulators (e.g. ETV5 and SPRY2), confirming their association with inhibitor sensitivity through established roles in pathway feedback and resistance. The approach also identified GKAP1 as a potential novel biomarker. For the BTK/MNK dual inhibitor, the identification of BCL2 and RASAL3 suggested dependencies and crosstalk with the RAS/ERK pathway. These findings underscore the robust biological relevance of our results, even when derived from modestly sized panels, and validate the utility of DRP-guided selection for identifying cell lines that reveal biologically interpretable mechanisms.

To address the critical need for early tumor selectivity screening, we applied the DRP methodology to compound selection, a common task in early-stage pharmaceutical development. Its effectiveness was confirmed with experimental validation, identifying compounds that selectively target tumorigenic (MCF7) over non-tumorigenic (MCF10A) breast epithelial cell lines. The successful *in silico* identification of Pevonedistat and BI 2536 as highly selective agents, coupled with the accurate classification of the two negative control compounds, demonstrates the method's potential to prioritise high-value candidates from compound libraries.

Although the DRP framework demonstrated robust performance, our findings also reveal important methodological considerations. The high variability in RMSE metrics between drugs (Table 3) indicates that predictive success strongly depends on the quality of the underlying DRP features. This reflects the inherent dependence of model performance on the composition of both the drug and cell line panels. To address this, future work should incorporate targeted experiments across more diverse datasets to elucidate the determinants of predictive success, enabling the refinement of panel composition for different use cases. Ultimately, adopting this DRP-based strategy represents a paradigm shift toward functional pharmacogenomics, offering a robust, cost-effective platform to accelerate mechanism-informed drug discovery and clinical translation.

**Conclusion**
This study introduces a validated, practical framework for accelerating early-stage drug development through Drug Response Profile (DRP) descriptors. By providing an orthogonal, functionally agnostic feature set, the DRP approach robustly supports critical decisions in discovery pipelines, ranging from mechanistic biomarker discovery to the efficient prioritization of selective drug candidates. Ultimately, this framework offers a cost-effective platform to accelerate the identification of promising compounds and promote the development of targeted therapies with higher success rates.

**Acknowledgments**

None of the above work would have been possible without the availability of cancer-derived cell lines. We would like to express gratitude to all patients who have made this, as well as many other studies contributing to the understanding and treatment of cancer, a possibility. This work has been supported by grants from the UK Engineering and Physical Sciences Research Council (EPSRC) [EP/R022925/2, EP/W004801/1 and EP/X032418/1].

**Data Availability**

GDSC 1 and 2 drug response data ($IC_{50}$ values - https://www.cancerrxgene.org/), CCLE drug response data ($IC_{50}$ values - https://depmap.org/), mRNA expression profiles (TPM values - https://cellmodelpassports.sanger.ac.uk/).

**Code Availability:**

Code and processed data is available in the github repository: https://github.com/abbiAR/-Strategic-Cell-Line-and-Compound-Selection-Using-Drug-Response-Profiles

**Competing interests**

E.T. works with Arctoris Ltd., a CRO in the area of drug development. The remaining authors declare no competing interests.

**Author Contributions**

A.A. designed the study, analysed the data, and wrote the main manuscript text. E.T. performed laboratory experiments and analysed the data. R.K. and L.S. supervised the study. All authors reviewed the work and the manuscript.